\documentclass{article}
\usepackage{amsfonts}

\usepackage{graphicx}
\usepackage{amsmath}

\newenvironment{proof}[1][Proof]{\textbf{#1.} }{\ \rule{0.5em}{0.5em}}
\input{tcilatex}

\begin{document}

\author{John Gough, Andrei Sobolev \\
%EndAName
Department of Computing \& Mathematics\\
Nottingham-Trent University, Burton Street,\\
Nottingham NG1\ 4BU, United Kingdom.\\
john.gough@ntu.ac.uk, andrei.sobolev@ntu.ac.uk}
\title{Quantum Markovian Approximations for Fermionic Reservoirs}
\date{}
\maketitle

\begin{abstract}
We establish a quantum functional central limit for the dynamics of a system
coupled to a Fermionic bath with a general interaction \ linear in the
creation, annihilation and scattering of the bath reservoir. Following a
quantum Markovian limit, we realize the open dynamical evolution of the
system as an adapted quantum stochastic process driven by Fermionic Noise.
\end{abstract}

\section{Introduction \& Statement Results}

\enlargethispage{\baselineskip}

Models of quantum systems driven by Fermionic Wiener processes have been
considered in \cite{Applebaum-Hudson}, \cite{Applebaum}. We extend the
results of \cite{Gough1} concerning quantum Markov approximations to the
case of a Fermi reservoir. The main technical difference comes about from
the absence of coherent vectors in Fermi Fock space and, of course, now we
have to control the sign which arises from the canonical anti-commutation
relations (CAR). This in turn extends a result of \cite{ACFRLU} on the weak
coupling limit.

We consider the evolution on a joint Hilbert space $\frak{H}_{S}\otimes 
\frak{H}_{R}$ governed by a time-dependent Hamiltonian $\Upsilon _{t}\left(
\lambda \right) =\sum_{\alpha ,\beta \left\{ \in 0,1\right\} }E_{\alpha
\beta }\otimes \left[ a_{t}^{+}\left( \lambda \right) \right] ^{\alpha }%
\left[ a_{t}^{-}\left( \lambda \right) \right] ^{\beta }$ where the fields $%
a_{t}^{\pm }\left( \lambda \right) $ are Fermionic fields on $\frak{H}_{R}$
and have an increasingly singular correlation as the parameter $\lambda
\rightarrow 0$. That is $\lim_{\lambda \rightarrow 0}\left\{ a_{s}^{-}\left(
\lambda \right) ,a_{t}^{+}\left( \lambda \right) \right\} =\gamma \delta
\left( s-t\right) $ for some $\gamma >0$.

We show that the unitary family $U_{t}\left( \lambda \right) :=\mathbf{\vec{T%
}}\exp \left\{ -i\int_{0}^{t}ds\,\Upsilon _{s}\left( \lambda \right)
\right\} $ converges in weak matrix elements to a unitary adapted quantum
stochastic process $U_{t}$ satisfying the quantum stochastic differential
equation driven by Fermionic Wiener processes $\mathbb{A}_{t}^{\pm }$ and by
the gauge process $\Lambda _{t}$ \cite{Applebaum-Hudson}, \cite{Applebaum}, 
\cite{HP:UFBSC}: 
\begin{equation}
dU_{t}=\frac{1}{\gamma }\left( W-1\right) U_{t}\otimes d\Lambda
_{t}+LU_{t}\otimes d\mathbb{A}_{t}^{+}-L^{\dagger }WU_{t}\otimes d\mathbb{A}%
_{t}^{-}-KU_{t}\otimes dt  \tag{1.1}
\end{equation}
where $W=\frac{1-i\kappa ^{\ast }E_{11}}{1+i\kappa E_{11}}\;$(unitary), $%
L=-i(1+i\kappa E_{11})^{-1}E_{10}\;$(bounded), and $K=\frac{1}{2}\gamma
L^{\dagger }L+iH$ with $H=E_{00}+ \text{Im}\kappa E_{01}\frac{1}{1+i\kappa
E_{11}}E_{10}\;$(self-adjoint). Here $\kappa $ is a complex damping constant
having a microscopic origin with $\gamma =2 \text{Re}\kappa $.

We fix some notation. Let $\Gamma \left( \frak{h}\right)
:=\bigoplus_{n=0}^{\infty }\frak{h}^{\otimes n}$ be the ``Full'' Fock space
over a fixed complex separable Hilbert space $\frak{h}$. The
(anti)-symmetrization operators $\frak{P}_{\pm }$ are defined through linear
extension of the relations $\frak{P}_{\pm }f_{1}\otimes \cdots \otimes f_{n}$
$:=$ $\frac{1}{n!}\sum_{\sigma \in \frak{S}_{n}}\left( \pm 1\right) ^{\sigma
} $ $f_{\sigma \left( 1\right) }\otimes \cdots \otimes f_{\sigma \left(
n\right) }$, with $f_{j}\in \frak{h}$, $\frak{S}_{n}$ denotes the
permutation group on $\left\{ 1,\dots ,n\right\} $ and $\left( -1\right)
^{\sigma }$ is the parity of the permutation $\sigma $. The Bose Fock space $%
\Gamma _{+}\left( \frak{h}\right) $ and the Fermi Fock space $\Gamma
_{-}\left( \frak{h}\right) $\ having $\frak{h}$ as one-particle space are
then defined as the subspaces $\Gamma _{\pm }\left( \frak{h}\right) :=\frak{P%
}_{\pm }\,\Gamma \left( \frak{h}\right) $. As usual, we distinguish the Fock
vacuum $\Phi =\left( 1,0,0\dots \right) $ which is common to both Bose and
Fermi spaces.

Let $h\in \frak{h}$, $U$ unitary and $H$ self-adjoint on $\frak{h}$. We
define the following operators on the Full Fock space 
\begin{eqnarray*}
C^{+}\left( h\right) \,f_{1}\otimes \cdots \otimes f_{n} &:&=\sqrt{n+1}%
\;h\otimes f_{1}\otimes \cdots \otimes f_{n}; \\
C^{-}\left( h\right) \,f_{1}\otimes \cdots \otimes f_{n} &:&=\frac{1}{\sqrt{n%
}}\;\left\langle h|f_{1}\right\rangle \,f_{2}\otimes \cdots \otimes f_{n}; \\
\Gamma \left( U\right) \,f_{1}\otimes \cdots \otimes f_{n} &:&=\left(
Uf_{1}\right) \otimes \cdots \otimes \left( Uf_{n}\right) ; \\
\gamma \left( H\right) \,f_{1}\otimes \cdots \otimes f_{n}
&:&=\sum_{j}f_{1}\otimes \cdots \otimes \left( Hf_{j}\right) \otimes \cdots
\otimes f_{n}.
\end{eqnarray*}

\enlargethispage{\baselineskip}

Bose creation and annihilation fields are then defined on $\Gamma _{+}\left( 
\frak{h}\right) $ as $B^{\pm }\left( h\right) :=\frak{P}_{+}\,C^{\pm }\left(
h\right) \,\frak{P}_{+}$ while Fermi creation and annihilation fields are
defined on $\Gamma _{-}\left( \frak{h}\right) $ as $A^{\pm }\left( h\right)
:=\frak{P}_{-}\,C^{\pm }\left( h\right) \,\frak{P}_{-}.$Using the
traditional conventions $\left[ X,Y\right] =XY-YX$ and $\left\{ X,Y\right\}
=XY+YX$, we have the canonical (anti)-commutation relations 
\begin{equation}
\text{CCR: } \left[ B^{-}\left( f\right) ,B^{+}\left( g\right) \right]
=\left\langle f|g\right\rangle;\quad \text{CAR: }\left\{ A^{-}\left(
f\right) ,A^{+}\left( g\right) \right\} =\left\langle f|g\right\rangle. 
\tag{1.2}
\end{equation}
Second quantization operators are defined as $\Gamma _{\pm }\left( U\right)
:=\frak{P}_{\pm }\,\Gamma \left( U\right) \,\frak{P}_{\pm }$ and
differential second quantization operators as $\gamma _{\pm }\left( H\right)
:=\frak{P}_{\pm }\,\gamma \left( H\right) \,\frak{P}_{\pm }$. We have the
relation $\exp \left\{ it\,\gamma _{\pm }\left( H\right) \right\} =\Gamma
_{\pm }\left( e^{itH}\right) $. More generally, we may take the argument of
the differential second quantizations to be bounded: for the rank-one
operator $H=\left| f\right\rangle \left\langle g\right| $, described in
standard Dirac bra-ket notation, we have $\gamma _{+}\left( \left|
f\right\rangle \left\langle g\right| \right) \equiv B^{+}\left( f\right)
B^{-}\left( g\right) $ and $\gamma _{-}\left( \left| f\right\rangle
\left\langle g\right| \right) \equiv A^{+}\left( f\right) A^{-}\left(
g\right) $.

\looseness = -1

In the Bose case, the exponential vector map $\varepsilon :\frak{h}\mapsto
\Gamma _{+}\left( \frak{h}\right) $ is introduced as $\varepsilon \left(
f\right) =\oplus _{n=0}^{\infty }\frac{1}{\sqrt{n!}}f^{\otimes n}$with $%
f^{\otimes n}$ the $n$-fold tensor product of $f$ with itself. The Fock
vacuum is, in particular, given by $\Phi =\varepsilon \left( 0\right) $. The
exponential vectors are frequently used in the analysis of second quantized
Boson fields and facilitate enormously dealings with creation, annihilation
and conservation operators. $\varepsilon \left( \frak{h}\right) $ is a total
subset in $\Gamma _{+}\left( \frak{h}\right) $. There is\ no analogue in the
Fermi case and here we have to treat $n$-particle vectors $\frak{P}%
_{-}f_{1}\otimes \cdots \otimes f_{n}$, sometimes written in exterior
algebra notation as $f_{1}\wedge \cdots \wedge f_{n}$, as a natural domain
for investigations. The operations of Bose and Fermi second quantization
have the natural functorial property $\Gamma _{\pm }\left( \frak{h}%
_{1}\oplus \frak{h}_{2}\right) \cong \Gamma _{\pm }\left( \frak{h}%
_{1}\right) \otimes \Gamma _{\pm }\left( \frak{h}_{2}\right) $, see \cite
{Partha}.

\section{The Model}

We shall consider a quantum mechanical system $S$ with a state space $\frak{H%
}_{S}$ coupled to a quantum Fermi field reservoir $R$ with a state space $%
\frak{H}_{R}=\Gamma _{-}(\frak{h}_{R}^{1})$. The interaction between the
system and reservoir will be given by the formal Hamiltonian 
\begin{equation*}
H^{(\lambda)}=H_{S}\otimes I_{R}+I_{S}\otimes H_{R}+\hbar
H_{Int}^{(\lambda)}.
\end{equation*}

The interaction Hamiltonian $H_{Int}^{(\lambda)}$ is described by 
\begin{equation}
H_{Int}^{(\lambda)}=E_{11}\otimes A^{+}(g)A^{-}(g)+\lambda E_{10}\otimes
A^{+}(g)+\lambda E_{01}\otimes A^{-}(g)+\lambda ^{2}E_{00}\otimes I_{R}, 
\tag{2.1}  \label{int_ham}
\end{equation}
where $E_{\alpha \beta }$ are bounded operators on $\,\frak{H}_{S}$ and $%
E_{\alpha \beta }^{\ast }=E_{\beta \alpha }$, $A^{\pm }$ are the Fermi
creation/annihilation operators which lead to a representation of the CAR
algebra over the Hilbert space $\frak{h}_{R}^{1}$.

\enlargethispage{\baselineskip}

We shall assume the following harmonic relations 
\begin{align*}
e^{-\tau H_{S}/i\hbar }E_{\alpha \beta }e^{\tau H_{S}/i\hbar }& =e^{i\omega
\tau (\beta -\alpha )}E_{\alpha \beta }, \\
e^{-\tau H_{S}/i\hbar }A^{\pm }(g)e^{\tau H_{S}/i\hbar }& =A^{\pm }(S_{\tau
}g).
\end{align*}
The family $S_{\tau }$ will be a one-parameter group of unitaries on $\,%
\frak{h}_{R}^{1}$ (operator $H_{R}$ is a differential second quantization of
the Stone generator $H_{R}^{1}$ of this family). We remove the free dynamics
by introducing the operator 
\begin{equation}
U(\tau ,\lambda )=e^{-\tau (H_{S}\otimes I_{R}+I_{S}\otimes H_{R})/i\hbar
}e^{\tau H^{(\lambda )}/i}.  \tag{2.2}  \label{U_def}
\end{equation}
Operator $U$ transforms to the interaction picture and satisfies the
equation 
\begin{equation}
\frac{\partial }{\partial t}U(t/\lambda ^{2},\lambda )=-i\Upsilon
_{t}(\lambda )U(t/\lambda ^{2},\lambda )  \tag{2.3}  \label{U_equ}
\end{equation}
where $\Upsilon _{t}(\lambda )=E_{\alpha \beta }\otimes \lbrack
a_{t}^{+}(\lambda )]^{\alpha }[a_{t}^{-}(\lambda )]^{\beta }$ and $%
a_{t}^{\pm }(\lambda )=\frac{1}{\lambda }A^{\pm }(S_{t/\lambda ^{2}}^{\omega
}g)$, $S_{\tau }^{\omega }=e^{-i\tau \omega }S_{\tau }$. (We employ the
summation convention that when the Greek indices $\alpha ,\beta ,\dots $ are
repeated then we sum each index over the values $0$ and $1$ - moreover we
understand the index $\alpha $ in $\left[ .\right] ^{\alpha }$ to represent
a power.) The operators $a_{t}^{\pm }(\lambda )$ have the following
correlation in the Fock vacuum state for the reservoir: 
\begin{equation}
\langle a_{t}^{-}(\lambda )a^{+}(\lambda )\rangle _{R}=G_{\lambda }(t-s):=%
\frac{1}{\lambda ^{2}}\langle g\;|\;S_{(t-s)/\lambda ^{2}}^{\omega }g\rangle
.  \tag{2.4}
\end{equation}

\section{The Wick ordering}

Our main tool will be the Wick ordering of certain operator products. To
simplify the calculation we need a special technique which we introduce in
this section. Let $\frak{h}$ be an arbitrary Hilbert space.

For any $\underline{f}=\left( f_{1},\ldots ,f_{n}\right) ,\underline{g}%
=\left( g_{1},\ldots ,g_{n}\right) \in \frak{h}^{\times n}$, and $\underline{%
\alpha },\underline{\beta }\in \{0,1\}^{n}$, we consider the product 
\begin{equation}
F_{n}\left( \underline{f},\underline{g};\underline{\alpha },\underline{\beta 
}\right) =[A^{+}(f_{n})]^{\alpha _{n}}[A^{-}(g_{n})]^{\beta _{n}}\ldots
\lbrack A^{+}(f_{1})]^{\alpha _{1}}[A^{-}(g_{1})]^{\beta _{1}},  \tag{3.1}
\label{WO_1}
\end{equation}
where $A^{\pm }$ are the Fermi creation/annihilation operators satisfying
the CAR over the Hilbert space $\frak{h}$. We have the relations 
\begin{equation}
\lbrack A^{-}(g_{i})]^{\beta _{i}}[A^{+}(f_{k})]^{\alpha _{k}}=\left(
-1\right) ^{\beta _{i}\alpha _{k}}[A^{+}(f_{k})]^{\alpha
_{k}}[A^{-}(g_{i})]^{\beta _{i}}+\beta _{i}\alpha _{k}\left\langle
g_{i}|f_{k}\right\rangle.  \tag{3.2}
\end{equation}

\smallskip

\noindent \textbf{Definition 3.1:} Let $S_{n}=\{1,\ldots ,n\}$; we denote by 
$\frak{G}_{n}$ the class of all partitions (that is, mutually exclusive,
collectively exhaustive collections of subsets) of $S_{n}$ and let $\frak{G}%
=\cup _{n=1}^{\infty }\frak{G}_{n}$. For $\mathcal{G}= \left\{ G_{1},\dots
,G_{m}\right\} \in \frak{G}_{n}$ we call each $G_{k}\subset S_{n}$ a part of
the partition $\mathcal{G}$: we then set $\mathcal{G}_{out}:=\left\{ \max
G_{k}:k=1,\dots ,m\right\} $ and $\mathcal{G}_{in}:=\left\{ \min
G_{k}:k=1,\dots ,m\right\} $. For $\underline{\alpha },\underline{\beta }\in
\{0,1\}^{n}$, we define the sign function $\xi $ on $\frak{G}_{n}$ by 
\begin{equation}
\xi (\mathcal{G},\alpha ,\beta ):=\sum\nolimits_{j>i}^{\prime }\beta
_{j}\alpha _{i}  \tag{3.3}
\end{equation}
where the sum is over all pairs $\left( j,i\right) $ such that $j>i$ and any
of the following conditions hold:

1. there exists a $i^{\prime }$ in the same part as $i$ and a $j^{\prime }$
in the same part as $j$ with $i^{\prime }>j>i>j^{\prime}$;

2. $i \in \mathcal{G}_{out}$ and there exists a $j^{\prime }$ in the same
part as $j$ with $j>i>j^{\prime}$;

2. $j \in \mathcal{G}_{in}$ and there exists a $i^{\prime }$ in the same
part as $i$ with $i^{\prime }>j>i$;

4. $i \in \mathcal{G}_{out}$ and $j \in \mathcal{G}_{in}$.

\smallskip \enlargethispage{\baselineskip}

Now we are going to introduce the Feynman type diagrams related to product (%
\ref{WO_1}); they will facilitate the explanation of the Wick ordering of
product (\ref{WO_1}). We arrange the elements of $S_{n}$ as vertices along
the line in descending order (figure 1 gives a typical example). At each
vertex $i\in S_{n}$ we have two lines: one representing a annihilation of $%
\beta _{i}=0,1$ reservoir quanta coming in from the right and one
representing a creation of $\alpha _{i}=0,1$ reservoir quanta going out to
the left. We choose pairs $(i_{1},j_{1}),\ldots ,(i_{k},j_{k})$ which
satisfy the conditions $i_{h}<j_{h}$, $i_{h}\neq i_{l}$, $j_{h}\neq j_{l}$,
where $h,l=1,\ldots ,k$ and $h\neq l$. Clearly, the number of pairs cannot
exceed $n-1$, but may be zero as well. For each pair $(i_{h},j_{h})$ we join
together (contract) the $\beta _{i_{h}}$ and $\alpha _{j_{h}}$ lines and get
the so called \textit{internal line} with label $\alpha _{j_{h}}\beta
_{i_{h}}$.

\begin{center}
\begin{picture}(240,42)
\label{pic1} \put(0,20){\line(1,0){240}} \put(40,20){\circle*{3}}
\put(80,20){\circle*{3}}\put(120,20){\circle*{3}}\put(160,20){\circle*{3}}
\put(200,20){\circle*{3}}\put(0,20){\oval(80,30)[tr]}
\put(80,20){\oval(80,30)[t]}\put(140,20){\oval(40,30)[t]}
\put(240,20){\oval(160,55)[tl]}\put(0,20){\oval(160,45)[tr]}
\put(140,20){\oval(120,45)[t]}\put(240,20){\oval(80,45)[tl]}
\put(35,12){$s_5$}\put(75,12){$s_4$}\put(115,12){$s_3$}\put(155,12){$s_2$}\put(195,12){$s_1$}
\put(28,25){$\alpha_5$}\put(45,25){$\beta_5$}
\put(68,25){$\alpha_4$}\put(85,25){$\beta_4$}
\put(108,25){$\alpha_3$}\put(125,25){$\beta_3$}
\put(148,25){$\alpha_2$}\put(165,25){$\beta_2$}
\put(188,25){$\alpha_1$}\put(205,25){$\beta_1$}\put(110,0){Figure
1}
\end{picture}
\end{center}

Each such diagram leads to an equivalence class on $S_{n}$ as follows: we
say that $x\equiv y$ if $x=y$ or if there exists a sequence of pairs
connecting $x$ to $y$. The equivalence classes give is a partition of $S_{n}$
and in this way every such diagram is uniquely associated with a partition
in $\frak{G}_{n}$. Indeed, each partition $\mathcal{G}\in \frak{G}_{n}$
gives a set of pairs $\cup _{\{i_{r}>\ldots >i_{1}\}\in \mathcal{G}%
,s>1}\{(i_{r},i_{r-1}),\ldots ,(i_{2},i_{1})\}$; the inverse correspondence
is obvious. Note that $\mathcal{G}_{in}$ is the set $\cup _{\{i_{r}>\ldots
>i_{1}\}\in \mathcal{G}}\{i_{1}\}$ and $\mathcal{G}_{out}$ is the set $\cup
_{\{i_{r}>\ldots >i_{1}\}\in \mathcal{G}}\{i_{s}\}$. The elements of set $%
\mathcal{G}_{in}$, $\mathcal{G}_{out}$ we shall call the vertices with 
\textit{external incoming, outgoing lines} respectively.

In the definition of the sign function $\xi $ we see that the sum is taken
over all line intersections, that is, the product $\beta _{j}\alpha _{i}$
includes in summation if $i<j$ and the $\beta _{j}$ line and $\alpha _{i}$
line intersect each other.

The diagram in figure 1 corresponds to the partition $\mathcal{G}%
=\{\{1,4\};\{2,3,5\}\}$;\ here function $\xi $ will be given by $\xi (%
\mathcal{G},\alpha ,\beta )=\beta _{2}\alpha _{1}+\beta _{4}\alpha
_{3}+\beta _{5}\alpha _{4}$; $\mathcal{G}_{out}=\{4,5\}$, $\mathcal{G}%
_{in}=\{1,2\}$.

\smallskip \enlargethispage{\baselineskip}

\noindent \textbf{Lemma 3.2:}\emph{\ }For any $\underline{f}=\left(
f_{1},\ldots ,f_{n}\right) ,\underline{g}=\left( g_{1},\ldots ,g_{n}\right)
\in \frak{h}^{\times n}$, and $\underline{\alpha },\underline{\beta }\in
\{0,1\}^{n}$, we have that the normal ordered form of $F_{n}\left( 
\underline{f},\underline{g};\underline{\alpha },\underline{\beta }\right) $
given in (3.1) is 
\begin{multline}
\sum_{\mathcal{G}\in \frak{G}_{n}}(-1)^{\xi (\mathcal{G},\underline{\alpha },%
\underline{\beta })}\prod_{\{i(r)>\ldots >i(1)\}\in \mathcal{G}}\left\{
\prod_{h=1}^{r-1}\beta _{i(h+1)}\alpha _{i(h)}\langle
g_{i(h+1)}|f_{i(h)}\rangle \right\}  \notag \\
\times \prod_{i\in \mathcal{G}_{out}}[A^{+}(f_{i})]^{\alpha _{i}}\prod_{j\in 
\mathcal{G}_{in}}[A^{-}(g_{j})]^{\beta _{j}}.  \tag{3.4}
\end{multline}

\begin{proof}
The proof is by induction on $n$. For $n=1$ there is nothing to prove so we
assume that $n>1$. We have that 
\begin{equation*}
F_{n+1}=[A^{+}(f_{n+1})]^{\alpha _{n+1}}[A^{-}(g_{n+1})]^{\beta _{n+1}}F_{n}
\end{equation*}
Using the induction hypothesis, we write $F_{n}$ as shown in (3.4). Let us
concentrate on the contribution from a fixed partition $\mathcal{G}\in \frak{%
G}_{n}$. We see that the annihilator $[A^{-}(g_{n+1})]^{\beta _{n+1}}$ will
be out of normal order and we use the relation (3.2) repeatedly to put it to
normal order. We first of all anti-commute it with the nearest creator,
which will be $[A^{+}(f_{i})]^{\alpha _{i}}$ where $i=\max \mathcal{G}_{out}$
and study the additional term involving the factor $\beta _{n+1}\alpha
_{i}\,\langle g_{n+1)}|f_{i}\rangle $. This is exactly the contribution to $%
F_{n+1}$ that should come from the partition $\mathcal{G}^{\prime }\in \frak{%
G}_{n+1}$ obtained by inserting the new vertex element $n+1$ into the part
of $\mathcal{G}$ containing $i$. Note that the sign function is unchanged
and this is consistent with the fact that no new intersections arise.

Proceeding to normal order in this manner we obtain all the new partitions
arising in this way with the correct multiplicative factor $\left( -1\right)
^{\beta _{n+1}\alpha _{i}}$, appearing at each stage. Eventually we are left
with then $n+1$ creators and annihilators put to normal order and this is
the contribution from the partition $\mathcal{G\cup }\left\{ n+1\right\} \in 
\frak{G}_{n+1}$: again the sign is consistent with the $n+1$ st. form of
(3.4). If we sum over all partitions $\mathcal{G}\in \frak{G}_{n}$ we see
that we have the appropriate contributions from all partitions in $\frak{G}%
_{n+1}$ as required.
\end{proof}

\bigskip

\noindent \textbf{Definition 3.3:}\emph{\ }Let $\mathcal{P}(S_{m},S_{n})$
denote the collection of all $P\subset S_{m}\times S_{n}$ such that if $%
(i_{1},j_{1}),(i_{2},j_{2})\in P$ then $i_{1}\neq i_{2}$ $%
\Longleftrightarrow $ $j_{1}\neq j_{2}$. For any $P\in \mathcal{P}%
(S_{m},S_{n})$, we denote by $P|_{S_{m}}$ the set $\{i\in S_{m}:\exists j\in
S_{n}$,$\text{ such that }(i,j)\in P\}$ and by $P|_{S_{n}}$ denote a set $%
\{j\in S_{n}:\exists i\in S_{m},\text{ such that }(i,j)\in P\}$: hopefully
no confusion should arise. We also set $\bar{P}|_{S_{m}}:=S_{m}\backslash
\,P|_{S_{m}}$ and $\bar{P}|_{S_{n}}:=S_{n}\backslash \,P|_{S_{n}}$.

\bigskip

\noindent \textbf{Lemma 3.4:}\emph{\ }We have the following normal ordering 
\begin{multline}
\prod_{j\in S_{n}}[A^{-}(g_{j})]^{\beta _{j}}\prod_{i\in
S_{m}}[A^{+}(f_{i})]^{\alpha _{i}}= \sum_{P\in \mathcal{P}%
(S_{m},S_{n})}(-1)^{\xi (P,\alpha ,\beta )} \\
\times \prod_{\left( i,j\right) \in P}\langle \beta _{j}g_{j}|\alpha
_{i}f_{i}\rangle \prod_{i\in \bar{P}|_{S_{m}}}[A^{+}(f_{i})]^{\alpha
_{i}}\prod_{j\in \bar{P}|_{S_{n}}}[A^{-}(g_{j})]^{\beta _{j}}.  \tag{3.5}
\label{WO_2}
\end{multline}

\smallskip

This is an immediate corollary to lemma 3.2. We set $\underline{\hat{f}}%
=(f_{1},\ldots ,f_{m},0,\dots ,0)$, $\underline{\hat{g}}=\left( 0,\dots
,0,g_{1},\ldots ,g_{n}\right) \in \frak{h}^{\times \left( n+m\right) }$ and $%
\underline{\hat{\alpha}}=(\alpha _{1},\ldots ,\alpha _{m},0,\ldots ,0),%
\underline{\hat{\beta}}=(0,\ldots ,0,$ $\beta _{1},\ldots ,\beta _{n})\in
\left\{ 0,1\right\} ^{n+m}$ then applying the general formula (\ref{WO_1})
to product (\ref{WO_2}) we can see that only those partitions of $S_{n+m}$
consisting of singletons and pairs as parts give a non-zero contribution to
the normal order formula, the creator element of pair has to belong to the
first $m$ indexes and the annihilator one has to belong to the last $n$
indexes. Clearly, the set of such partitions has one-to-one relation with
the set $\mathcal{P}(S_{m},S_{n})$. We have used the notation $\xi $ for the
map whose exact definition is $\xi (P,\alpha ,\beta ):=\xi (\mathcal{G}_{P},%
\underline{\hat{\alpha}},\underline{\hat{\beta}})$, where $\underline{\hat{%
\alpha}}$ and $\underline{\hat{\beta}}$ were constructed as above and $%
\mathcal{G}_{P}$ is a partition of $S_{n+m}$ corresponded to the set of
pairs $P$.

\section{The Dyson series expansion}

\enlargethispage{\baselineskip}

By $\frak{k}$ denote the maximal subspace of $\,\frak{h}_{R}^{1}$ which $\,$%
satisfies the following condition: $\int_{\mathbb{R}}|\langle
f_{1}\;|\;S_{u}^{\omega }f_{2}\rangle |du<\infty $ whenever $f_{1},f_{2}\in 
\frak{k}$. The sesquilinear form on $\frak{k}$ is defined by $%
(f_{1}|f_{2}):=\int_{\mathbb{R}}\langle f_{1}\;|\;S_{u}^{\omega
}f_{2}\rangle du$. We also denote by $\frak{k}$ the associated Hilbert
space, i.e. the completion of the quotient of $\frak{k}$ by zero $%
(\,\cdot\,|\,\cdot\,)$-norm elements. We shall suppose below that the test
vector $g$ appearing in the interaction Hamiltonian (\ref{int_ham}) belongs
to $\frak{k}$. Define 
\begin{equation*}
\kappa =(g|g)_{+}:=\int_{0}^{\infty }G_{1}(u)du,\quad K:=\int_{0}^{\infty
}|G_{1}(u)|du
\end{equation*}
then we have the Markovian limit for the two-point functions 
\begin{equation}
\lim_{\lambda \rightarrow 0}G_{\lambda }(t-s)=\kappa \frak{d}%
_{+}(t-s)+\kappa ^{\ast }\frak{d}_{-}(t-s),  \tag{4.1}  \label{G_lim}
\end{equation}
where $\frak{d}_{\pm }$ are generalized functions defined by (see \cite
{Gough1}) 
\begin{equation}
\langle \frak{d}_{\pm },f\rangle :=\int \frak{d}_{\pm }(t-s)f(s)ds=f(t^{\pm
}).  \tag{4.2}
\end{equation}

For any $f\in \frak{k}$ and $S,T,\lambda \in \mathbb{R}$ define a collective
vector $\tilde{f}$ by the rule 
\begin{equation}
\tilde{f}(\lambda ):=\frac{1}{\lambda }\int_{S}^{T}du\,S_{u/\lambda
^{2}}^{\omega }f.  \tag{4.3}  \label{col_vect}
\end{equation}

Let us fix vectors $\varphi _{1},\varphi _{2}\in \,\frak{H}_{S}$, integers $%
k_{+},k_{-}\in \mathbb{N}_{0}$, vectors $f_{1}^{+},\ldots ,f_{k_{+}}^{+},%
\newline
f_{1}^{-},\ldots ,f_{k_{-}}^{-}\in \frak{k}$ and numbers $%
T_{1}^{+},S_{1}^{+},\ldots
,T_{k_{+}}^{+},S_{k_{+}}^{+},T_{1}^{-},S_{1}^{-},\ldots
,T_{k_{-}}^{-},S_{k_{-}}^{-}\in \mathbb{R}$ such that $T_{i}^{+}>S_{i}^{+}$
and $T_{i}^{-}>S_{i}^{-}$ for all $i$. To each quadruple $(f_{i}^{\pm
},S_{i}^{\pm },T_{i}^{\pm },\lambda )$ we put into the correspondence the
collective vector $\tilde{f}_{i}^{\pm }(\lambda )$ defined by (\ref{col_vect}%
). Denote 
\begin{equation}
h_{i}^{\pm }(t,\lambda ):=\frac{1}{\lambda}\langle \tilde{f}_{i}^{\pm
}(\lambda )\;|\;S_{t/\lambda ^{2}}^{\omega }g\rangle =\frac{1}{\lambda ^{2}}%
\int_{S_{i}^{\pm }}^{T_{i}^{\pm }}du\langle S_{u/\lambda ^{2}}^{\omega
}f_{i}^{\pm }\;|\;S_{t/\lambda ^{2}}^{\omega }g\rangle .  \tag{4.4}
\label{h_def}
\end{equation}

We shall be interested in the behavior of inner product 
\begin{equation}
\langle \phi _{1}\otimes \underset{i\in S_{k_{-}}}{\prod\nolimits^{\prime }}%
A^{+}(\tilde{f}_{i}^{-}(\lambda ))\Phi \;|\;U(t/\lambda ^{2},\lambda )\phi
_{2}\otimes \prod_{i\in S_{k_{+}}}A^{+}(\tilde{f}_{i}^{+}(\lambda ))\Phi
\rangle  \tag{4.5}  \label{inner_product}
\end{equation}
in the limit $\lambda \rightarrow 0$ (the so called $\delta -$\textit{%
correlated noise limit}). Here and below we use sign $\prod\nolimits^{\prime
}$ to denote the product in inverse order. The operator $U(t/\lambda
^{2},\lambda )$ here, defined in (\ref{U_def}), can be developed as a formal
Dyson series: 
\begin{equation*}
U(t/\lambda ^{2},\lambda )=\sum_{n=0}^{\infty }(-i)^{n}D_{n}(t,\lambda )
\end{equation*}
where we have the multi-time integral 
\begin{equation*}
D_{n}(t,\lambda )=\int_{\triangle _{n}(t)}ds_{n}\ldots ds_{1}\;\Upsilon
_{s_{1}}(\lambda )\ldots \Upsilon _{s_{n}}(\lambda )
\end{equation*}
over the simplex $\triangle _{n}(t):=\{(s_{n},\ldots
,s_{1}):\;t>s_{n}>\ldots >s_{1}>0\}$.

\smallskip \enlargethispage{\baselineskip}

\noindent \textbf{Lemma 4.1: }For any $k_{+},k_{-}\in \mathbb{N}_{0}$, $%
f_{1}^{-},\ldots ,f_{k_{-}}^{-},f_{1}^{+},\ldots ,f_{k_{+}}^{+}\in \frak{k}$%
, $\{S_{i}^{-},T_{i}^{-}\}_{i=1}^{k_{-}}$, $\{S_{i}^{+},T_{i}^{+}%
\}_{i=1}^{k_{+}}\subset \mathbb{R}$, $t\geq 0$, we have 
\begin{multline}
\langle \varphi _{1}\otimes \underset{i\in S_{k_{-}}}{\prod\nolimits^{\prime
}}A^{+}(\tilde{f}_{i}^{-}(\lambda ))\Phi \;|\;D_{n}(t,\lambda )\varphi
_{2}\otimes \prod_{i\in S_{k_{+}}}A^{+}(\tilde{f}_{i}^{+}(\lambda ))\Phi
\rangle  \notag \\
=\sum_{\alpha ,\beta \in \{0,1\}^{n}}\langle \varphi _{1}\,|\,E_{\alpha
_{n}\beta _{n}}\ldots E_{\alpha _{1}\beta _{1}}\varphi _{2}\rangle \sum_{%
\mathcal{G}\in \frak{G}_{n}}\sum_{P_{1}\in \mathcal{P}(S_{k_{-}},\mathcal{G}%
_{out})}\sum_{P_{2}\in \mathcal{P}(S_{k_{+}},\mathcal{G}_{in})}  \notag \\
(-1)^{\xi (\mathcal{G},\alpha ,\beta )+\xi (P_{1},\alpha ,1)+\xi
(P_{2},1,\beta )}\delta _{0}\left( \sum_{i\in \bar{P_{1}}|_{\mathcal{G}%
_{out}}}\alpha _{i}+\sum_{i\in \bar{P_{2}}|_{\mathcal{G}_{in}}}\beta
_{i}\right)  \notag \\
\times \langle \underset{i\in \bar{P_{1}}|_{S_{k_{-}}}}{\prod\nolimits^{%
\prime }}A^{+}(\tilde{f}_{i}^{-}(\lambda ))\Phi \;|\;\prod_{i\in \bar{P_{2}}%
|_{S_{k_{+}}}}A^{+}(\tilde{f}_{i}^{+}(\lambda ))\Phi \rangle  \notag \\
\times \int_{\triangle _{n}(t)}ds_{n}\ldots ds_{1}\prod_{\{i(r)>\dots
>i(1)\}\in \mathcal{G}}\left\{ \prod_{h=1}^{r-1}\beta _{i(h+1)}\alpha
_{i(h)}G_{\lambda }(s_{i(h+1)}-s_{i(h)})\right\}  \notag \\
\times \prod_{(i_{1},j_{1})\in P_{1}}\alpha
_{j_{1}}h_{i_{1}}^{-}(s_{j_{1}},\lambda )\prod_{(i_{2},j_{2})\in
P_{2}}(\beta _{j_{2}}h_{i_{2}}^{+}(s_{j_{2}},\lambda ))^{\ast }  \tag{4.6}
\label{n_term}
\end{multline}

\begin{proof}
Note that 
\begin{multline}
\Upsilon _{s_{n}}(\lambda )\ldots \Upsilon _{s_{1}}(\lambda )=  \notag \\
\sum_{\alpha ,\beta \in \{0,1\}^{n}}E_{\alpha _{n}\beta _{n}}\ldots
E_{\alpha _{1}\beta _{1}}[a_{s_{n}}^{+}(\lambda )]^{\alpha
_{n}}[a_{s_{n}}^{-}(\lambda )]^{\beta _{n}}\ldots \lbrack
a_{s_{1}}^{+}(\lambda )]^{\alpha _{1}}[a_{s_{1}}^{-}(\lambda )]^{\beta _{1}}.
\tag{4.7}  \label{equ1}
\end{multline}
The proof consists of a direct application of lemma 3.2 to (\ref{equ1}) and
then we apply twice the formula (\ref{WO_2}). Note that we do not get a Wick
ordering; each additional term, arising from an application of the CAR, will
include a product of operators of the form 
\begin{equation*}
\prod_{i\in \bar{P_{1}}|_{\mathcal{G}_{out}}}[a_{s_{i}}^{+}(\lambda
)]^{\alpha _{i}}\prod_{i\in \bar{P_{1}}|_{S_{k_{-}}}}A^{-}(\tilde{f}%
_{i}^{-}(\lambda ))\prod_{i\in \bar{P_{2}}|_{S_{k_{+}}}}A^{+}(\tilde{f}%
_{i}^{+}(\lambda ))\prod_{i\in \bar{P_{2}}|_{\mathcal{G}%
_{in}}}[a_{s_{i}}^{-}(\lambda )]^{\beta _{i}}.
\end{equation*}
Taking the vacuum expectation we get zero whenever any $\alpha $ or $\beta $
from the correspondent sets equals to unity. So using the Kronecker delta $%
\delta _{0}\left( .\right) $ we can write the vacuum expectation in the
following form 
\begin{equation*}
\delta _{0}\left( \sum_{i\in \bar{P_{1}}|_{\mathcal{G}_{out}}}\alpha
_{i}+\sum_{i\in \bar{P_{2}}|_{\mathcal{G}_{in}}}\beta _{i}\right) \langle 
\underset{i\in \bar{P_{1}}|_{S_{k_{-}}}}{\prod\nolimits^{\prime }}A^{+}(%
\tilde{f}_{i}^{-}(\lambda ))\Phi \;|\;\prod_{i\in \bar{P_{2}}%
|_{S_{k_{+}}}}A^{+}(\tilde{f}_{i}^{+}(\lambda ))\Phi \rangle .
\end{equation*}
\end{proof}

\bigskip

\noindent \textbf{Definition 4.2: }The partition $\mathcal{G}\in \frak{G}%
_{n} $ is called the type $I$ partition if each subset $\{i(r)>\ldots
>i(1)\}\in \mathcal{G}$ $\,$satisfies the condition $i(2)-i(1)=i(3)-i(2)=%
\ldots =i(r)-i(r-1)=1$. We shall denote by $\frak{G}_{I,n}$ a set of all
type $I$ partitions. The rest of partitions forms a set $\frak{G}_{II,n}$,
they are called the type $II$ partitions. Set also $\frak{G}_I:=\cup _{n}%
\frak{G}_{I,n}$. Given $\mathcal{G}\in \frak{G}$ we denote the number of
vertices partitioned by $\mathcal{G}$ as $E\left( \mathcal{G}\right) $ and
the number of parts making up $\mathcal{G}$ as $N\left( \mathcal{G}\right) $.

\smallskip

Note that type $I$ partitions correspond to diagrams where the contractions
are between \emph{pairs of consecutive vertices} only.

\smallskip

\noindent \textbf{Lemma 4.3:}\label{type_two} For any $\mathcal{G}\in \frak{G%
}_{II,n}$, $k_{+},k_{-}\in \mathbb{N}_{0}$, $f_{1}^{-},\ldots
,f_{k_{-}}^{-},f_{1}^{+},\ldots ,f_{k_{+}}^{+}\in \frak{k}$, $%
\{S_{i}^{-},T_{i}^{-}\}_{i=1}^{k_{-}},\{S_{i}^{+},T_{i}^{+}\}_{i=1}^{k_{+}}%
\subset \mathbb{R}$, $t\geq 0$, we have 
\begin{gather}
\lim_{\lambda \rightarrow 0}\int_{\triangle _{n}(t)}ds_{n}\ldots
ds_{1}\prod_{\{i(r)>\ldots >i(1)\}\in \mathcal{G}}\left\{
\prod_{h=1}^{r-1}\beta _{i(h+1)}\alpha _{i(h)}G_{\lambda
}(s_{i(h+1)}-s_{i(h)})\right\}  \notag \\
\times |\langle \Phi \;|\;\prod_{j\in S_{k_{-}}}A^{-}(\tilde{f}%
_{j}^{-}(\lambda ))\prod_{i\in \mathcal{G}_{out}}[a_{s_{i}}^{+}(\lambda
)]^{\alpha _{i}} \prod_{j\in \mathcal{G}_{in}}[a_{s_{j}}^{-}(\lambda
)]^{\beta _{j}}\prod_{i\in S_{k_{+}}}A^{+}(\tilde{f}_{i}^{+}(\lambda))\Phi
\rangle |=0.  \tag{4.8}
\end{gather}

\begin{proof}
The inner product in (4.8) is bounded for all $\lambda >0$. The proof of the
identity 
\begin{equation*}
\lim_{\lambda \rightarrow 0} \int_{\triangle _{n}(t)}ds_{n}\ldots
ds_{1}\prod_{\{i(r)>\ldots >i(1)\}\in \mathcal{G}}\left\{
\prod_{h=1}^{r-1}G_{\lambda }(s_{i(h+1)}-s_{i(h)})\right\} =0
\end{equation*}
for $\mathcal{G}\in \frak{G}_{II,n}$ repeats the proof of \cite[Lemma 6.1]
{Gough1}.
\end{proof}

\bigskip

\noindent \textbf{Lemma 4.4:}\label{limits} For any $\tilde{f}%
_{i}^{+}(\lambda)$ and $\tilde{f}_{j}^{-}(\lambda)$ we have 
\begin{equation}
\lim_{\lambda \rightarrow 0}\langle \tilde{f}_{j}^{-}(\lambda)\,|\,\tilde{f}%
_{i}^{+}(\lambda)\rangle =\langle \chi _{\lbrack S_{j}^{-},T_{j}^{-}]},\chi
_{\lbrack S_{i}^{+},T_{i}^{+}]}\rangle (f_{j}^{-}|f_{i}^{+}).  \tag{4.9}
\label{f_lim}
\end{equation}
2. The functions $h_{i}^{\pm }(t,\lambda )$ defined in (\ref{h_def}) will
have the limits 
\begin{equation}
\lim_{\lambda \rightarrow 0}h_{i}^{\pm }(t,\lambda )=h_{i}^{\pm
}(t):=1_{[S_{i}^{\pm },T_{i}^{\pm }]}(t)(f_{i}^{\pm }|g).  \tag{4.10}
\label{h_lim}
\end{equation}

For the proof of first statement see \cite[Lemma 3.2]{ACFRLU}. The second
statement can be proved similarly.

\section{Uniform convergence of Dyson series}

With each partition $\mathcal{G}\in \frak{G}_{n}$ we associate a sequence of
occupation numbers $\mathbf{n}=(n_{j})_{j=1}^{\infty }$ where $%
n_{j}=0,1,2,\ldots $ counts the number of $j$-tuples making up $\mathcal{G}$
(see \cite{Gough1}). We put, by definition 
\begin{equation}
E(\mathbf{n}):=\sum_{j}jn_{j},\quad N(\mathbf{n}):=\sum_{j}n_{j}.  \tag{5.1}
\end{equation}
Denote by $\mathcal{G}_{0}(\mathbf{n})$ a partition where we have all $1$%
-tuples in the beginning, then followed by all $2$-tuples, etc. A
permutation $\rho $ of set $S_{E(\mathbf{n})}$ is called admissible if it
maps the partition $\mathcal{G}_{0}(\mathbf{n})$ into another partition $%
\rho (\mathcal{G}_{0}(\mathbf{n}))$. We shall denote by$\ \frak{S}_{\mathbf{n%
}}^{0}$ the collection of all admissible permutations $\rho $.

We have the following inequalities: 
\begin{align*}
{}& |h_{i}^{\pm }(t,\lambda )| \leq \int_{\mathbb{R}}|\langle S_{u}^{\omega
}f_{i}^{\pm }\,|\,g\rangle |du \; \text{ for all } \lambda>0,\;t \geq 0; \\
{}& |\langle\tilde{f}^-_i(\lambda)\;|\;\tilde{f}^+_j(\lambda)\rangle| \leq
(T^-_i - S^-_i) \int_{\mathbb{R}} |\langle f^-_i\;|\; S^{\omega}_u f^+_j
\rangle| du \; \text{ for all } \lambda>0.
\end{align*}
Denote $C_{h}:=\max \{1\,,\,\max_{i}\int_{\mathbb{R}}|\langle S_{u}^{\omega
}f_{i}^{\pm }\,|\,g\rangle |du \}$, $C_{f}:=\max \{1\,,\, \max_{i,j}(T^-_i -
S^-_i)\newline
\int_{\mathbb{R}} |\langle f^-_i\;|\; S^{\omega}_u f^+_j \rangle| du\}$, $%
C_{11}:=||E_{11}||$ and $C:=\max
\{||E_{11}||,||E_{10}||,||E_{01}||,||E_{00}||\}$.

Denote also 
\begin{equation}
I(n,\mathcal{G}):=\int_{\triangle _{n}(t)}ds_{n}\ldots
ds_{1}\prod_{\{i(r)>\ldots >i(1)\}\in \mathcal{G}}\left\{
\prod_{h=1}^{r-1}|G_{\lambda }(s_{i(h+1)}-s_{i(h)})|\right\}  \tag{5.2}
\end{equation}

\noindent \textbf{Theorem 5.1:}\label{uniconv} Suppose $K||E_{11}||<1$; then
the series 
\begin{equation}  \label{series1}
\sum_{n}(-i)^{n}\langle \underset{i\in S_{k^{-}}}{\prod\nolimits^{\prime }}%
A^{+}(\tilde{f}_{i}^{-}(\lambda))\Phi \;|\;D_{n}(t,\lambda )\prod_{i\in
S_{k^{+}}}A^{+}(\tilde{f}_{i}^{+}(\lambda))\Phi \rangle  \tag{5.3}
\end{equation}
converges uniformly and absolutely in the pair $(\lambda ,t)\in \mathbb{R}%
_{+}\times \lbrack 0,T]$ for any $T<\infty $.

\begin{proof}
We have to estimate the absolute value of right side of equality (\ref
{n_term}).

First, for given $\alpha $, $\beta $ and $\mathcal{G}$ we need to estimate
the maximal number of sets $P_{1}\in \mathcal{P}(S_{k_{-}},\mathcal{G}%
_{out}) $, which give non-zero contribution (we call them non-trivial sets).
Denote $|\mathcal{G}_{out}|:=\sum_{i\in \mathcal{G}_{out}}\alpha _{i}$.
Clearly, if we have $|\mathcal{G}_{out}|>k_{-}$ then all $P_{1}\in \mathcal{P%
}(S_{k_{-}},\mathcal{G}_{out})$ gives a zero contribution (it is provided by
function $\delta _{0}$). So we can consider only the case $|\mathcal{G}%
_{out}|\leq k_{-}$. We can also deduce that a non-trivial set $P_{1}$ cannot
include a pair $(i,j)$ such that $\alpha _{j}=0$. It means that the number
of non-trivial sets $P_{1}\in \mathcal{P}(S_{k_{-}},\mathcal{G}_{out})$
cannot exceed the cardinality of $\mathcal{P}(S_{k_{-}},S_{k_{-}})$ which is
equal to $\sum_{i=0}^{k_{-}}\left( \frac{(k_{-})!}{i!(k_{-}-i)!}\right)
^{2}i!\leq (k_{-})!2^{k_{-}}$. Similarly, we have the upper estimate $%
(k_{+})!2^{k_{+}}$ for the maximal number of non-trivial sets $P_{2}\in 
\mathcal{P}(S_{k_{+}},\mathcal{G}_{in})$.

We have the following estimates: 
\begin{align*}
{}& |\langle \underset{i\in \bar{P}_{1}|_{S_{k_{-}}}}{\prod\nolimits^{\prime
}}A^{+}(\tilde{f}_{i}^{-}(\lambda ))\Phi \;|\;\prod_{i\in \bar{P}%
_{2}|_{S_{k_{+}}}}A^{+}(\tilde{f}_{i}^{+}(\lambda ))\Phi \rangle | \leq
(k_+\wedge k_-)! (C_f)^{k_+\wedge k_-}, \\
{}& |\prod_{(i_{1},j_{1})\in P_{1}}\alpha
_{j_{1}}h_{i_{1}}^{-}(s_{j_{1}},\lambda )\prod_{(i_{2},j_{2})\in
P_{2}}(\beta _{j_{2}}h_{i_{2}}^{+}(s_{j_{2}},\lambda ))^{\ast }| \leq
(C_{h})^{k_{-}}(C_{h})^{k_{+}}.
\end{align*}

Hence, we obtain 
\begin{multline*}
|\langle \underset{i\in S_{k^{-}}}{\prod\nolimits^{\prime }}A^{+}(\tilde{f}%
_{i}^{-}(\lambda))\Phi \;|\;D_{n}(t,\lambda )\prod_{i\in S_{k^{+}}}A^{+}(%
\tilde{f}_{i}^{+}(\lambda))\Phi \rangle | \leq (k_+\wedge k_-)!
(C_f)^{k_+\wedge k_-} \\
\times (k_{-})!(k_{+})!(2C_{h})^{k_{-}+k_{+}}\sum_{\mathcal{G}\in \frak{G}%
_{n}}\underset{\alpha ,\beta \in \{0,1\}^{n}}{\sum\nolimits^{\prime }}
|\langle \varphi _{1}\,|\,E_{\alpha _{n}\beta _{n}}\ldots E_{\alpha
_{1}\beta _{1}}\varphi _{2}\rangle |I(n,\mathcal{G}) \\
=(k_+\wedge k_-)! (C_f)^{k_+\wedge k_-}(k_{-})!(k_{+})!(2C_{h})^{k_{-}+k_{+}}
\\
\times \sum_{\mathbf{n}}^{E(\mathbf{n})=n}\sum_{\rho \in \frak{S}_{\mathbf{n}%
}^{0}}I(n,\rho (\mathcal{G}(\mathbf{n})))\underset{\alpha ,\beta \in
\{0,1\}^{n}}{\sum\nolimits^{\prime }}|\langle \varphi _{1}\,|\,E_{\alpha
_{n}\beta _{n}}\ldots E_{\alpha _{1}\beta _{1}}\varphi _{2}\rangle |.
\end{multline*}
Here $\sum\nolimits^{\prime }$ means that we sum up only over those $\alpha $
and $\beta $ which give a non-zero contribution for given $\mathcal{G}\in 
\frak{G}_{n}$. Indeed, for $\mathcal{G}=\mathcal{G}(\mathbf{n})$ the number
of fixed variables $\alpha $ and $\beta $ is equal to $2(E(\mathbf{n})-N(%
\mathbf{n}))$ (they equal to unity). Note that the estimate of $|\langle
\varphi _{1}\,|\,E_{\alpha _{n}\beta _{n}}\ldots E_{\alpha _{1}\beta
_{1}}\varphi _{2}\rangle |$ depends only on $\mathbf{n}$. We can vary only $%
N(\mathbf{n})$ values of $\alpha $ and $N(\mathbf{n})$ values of $\beta $.
When we have all these unfixed variables equal to zero we obtain the upper
bound $C_{11}^{E(\mathbf{n})-2N(\mathbf{n})+n_{1}}C^{2N(\mathbf{n})-n_{1}}$.
When we set any variable equal to unity it means that we replace 1) an $%
E_{00}$ operator by $E_{01}$ or $E_{10}$, or 2) an $E_{10}$ or $E_{01} $
operator by $E_{11}$. In the first case the estimate stays the same; in the
second case we get the estimate which is a product of the previous one and a
factor $C_{11}C^{-1}$. Note that this factor does not exceed the unity so
the previous estimate is valid also. Thus, for given $\mathbf{n}$, we have
the estimate $|\langle \varphi _{1}\,|\,E_{\alpha _{n}\beta _{n}}\ldots
E_{\alpha _{1}\beta _{1}}\varphi _{2}\rangle |\leq C_{11}^{E(\mathbf{n})-2N(%
\mathbf{n})+n_{1}}C^{2N(\mathbf{n})-n_{1}}\leq C_{11}^{E(\mathbf{n})-2N(%
\mathbf{n})}C^{2N(\mathbf{n})}$, independent of the variables $\alpha $ and $%
\beta $. Clearly, we have $2^{2N(\mathbf{n})}$ possible values of $\alpha $
and $\beta $, so for given $\mathbf{n}$ the total estimate of sum $%
\sum^{\prime }$ is 
\begin{equation*}
\underset{\alpha ,\beta \in \{0,1\}^{n}}{\sum\nolimits^{\prime }}|\langle
\varphi _{1}\,|\,E_{\alpha _{n}\beta _{n}}\ldots E_{\alpha _{1}\beta
_{1}}\varphi _{2}\rangle |\leq C_{11}^{E(\mathbf{n})-2N(\mathbf{n})}(2C)^{2N(%
\mathbf{n})}.
\end{equation*}

We have obtained the following estimate for $n$-th term of series (\ref
{series1}) 
\begin{multline}
|\langle \underset{i\in S_{k^{-}}}{\prod\nolimits^{\prime }}A^{+}(\tilde{f}%
_{i}^{-}(\lambda))\Phi \;|\;D_{n}(t,\lambda )\prod_{i\in S_{k^{+}}}A^{+}(%
\tilde{f}_{i}^{+}(\lambda))\Phi \rangle |\leq (k_+\wedge k_-)!
(C_f)^{k_+\wedge k_-} \\
\times (k_{-})!(k_{+})! (2C_{h})^{k_{-}+k_{+}}\sum_{\mathbf{n}}^{E(\mathbf{n}%
)=n}C_{11}^{E(\mathbf{n})-2N(\mathbf{n})}(2C)^{2N(\mathbf{n})}\sum_{\rho \in 
\frak{S}_{\mathbf{n}}^{0}}I(n,\rho (\mathcal{G}(\mathbf{n}))).  \tag{5.4}
\label{conv}
\end{multline}

The uniform convergence in the pair $(\lambda ,t)\in \mathbb{R}_{+}\times
\lbrack 0,T]$ for any $T<\infty $ of series with $n$-th term given by right
side of inequality (\ref{conv}) under the condition $KC_{11}<1$ was proved
in \cite[Section 7]{Gough1}. It proves the absolute and uniform convergence
of series (\ref{series1}) under the same condition.
\end{proof}

\section{Limit QSDE}

Let $\mathbb{A}^{\pm }(\cdot )$ and $\mathbf{\Lambda }(\cdot )$ be the
Fermionic creation/annihilation and differential second quantization fields
with test functions in $\frak{k}\otimes L^{2}(\mathbb{R}^{+})$. Let $\Psi $
be a vacuum vector in $\Gamma _{-}(\frak{k}\otimes L^{2}(\mathbb{R}^{+}))$.

\bigskip

\noindent \textbf{Theorem 6.1:}\label{WCL} Suppose $K||E_{11}||<1$, then for
any $\varphi _{1},\varphi _{2}\in \,\frak{H}_{S}$, $k_{+},k_{-}\in \mathbb{N}%
_{0}$, $f_{1}^{-},\ldots ,f_{k_{-}}^{-},f_{1}^{+},\ldots ,f_{k_{+}}^{+}\in 
\frak{k}$, $\{S_{i}^{-},T_{i}^{-}\}_{i=1}^{k_{-}},\{S_{i}^{+},T_{i}^{+}%
\}_{i=1}^{k_{+}}\subset \mathbb{R}$, $t\geq 0$, we have 
\begin{multline}  \label{equ61}
\lim_{\lambda \rightarrow 0}\langle \phi _{1}\otimes \underset{i\in S_{k_{-}}%
}{\prod\nolimits^{\prime }}A^{+}(\tilde{f}_{i}^{-}(\lambda ))\Phi
\;|\;U(t/\lambda ^{2},\lambda )\phi _{2}\otimes \prod_{i\in S_{k_{+}}}A^{+}(%
\tilde{f}_{i}^{+}(\lambda ))\Phi \rangle  \notag \\
=\sum_{\mathcal{G}\in \frak{G}_{I}}\sum_{\alpha ,\beta \in \{0,1\}^{E(%
\mathcal{G})}}\langle \varphi _{1}\,|\,E_{\alpha _{E(\mathcal{G})}\beta _{E(%
\mathcal{G})}} \ldots E_{\alpha _{1}\beta _{1}} \varphi _{2}\rangle
\sum_{P_{1}\in \mathcal{P}(S_{k_{-}},\mathcal{G}_{out})}\sum_{P_{2}\in 
\mathcal{P}(S_{k_{+}},\mathcal{G}_{in})}  \notag \\
\times (-i)^{E(\mathcal{G})}(-1)^{\xi (\mathcal{G},\alpha ,\beta )+\xi
(P_{1},\alpha ,1)+\xi (P_{2},1,\beta )}\delta _{0}\left( \sum_{i\in \bar{%
P_{1}}|_{\mathcal{G}_{out}}}\alpha _{i}+\sum_{i\in \bar{P_{2}}|_{\mathcal{G}%
_{in}}}\beta _{i}\right)  \notag \\
\times \langle \underset{i\in \bar{P_{1}}|_{S_{k_{-}}}}{\prod\nolimits^{%
\prime }}\mathbb{A}^{+}(f_{i}^{-}\otimes 1_{ \lbrack
S_{i}^{-},T_{i}^{-}]})\Psi \;|\;\prod_{i\in \bar{P_{2}}|_{S_{k_{+}}}}\mathbb{%
A}^{+}(f_{i}^{+}\otimes 1_{[S_{i}^{+},T_{i}^{+}]})\Psi \rangle  \notag \\
\times \int_{\triangle _{n}(t)}ds_{E(\mathcal{G})}\ldots
ds_{1}\prod_{(i_{1},j_{1})\in P_{1}}\alpha
_{j_{1}}h_{i_{1}}^{-}(s_{j_{1}})\prod_{(i_{2},j_{2})\in P_{1}}(\beta
_{j_{2}}h_{i_{2}}^{+}(s_{j_{2}}))^{\ast }  \notag \\
\times \prod_{\{i(r)>\ldots >i(1)\}\in \mathcal{G}}\kappa ^{r-1}\left\{
\prod_{h=1}^{r-1}\beta _{i(h+1)}\alpha _{i(h)}\frak{d}%
_{+}(s_{i(h+1)}-s_{i(h)}) \right\}  \tag{6.1}
\end{multline}

\begin{proof}
By theorem 5.1 under the condition $K||E_{11}||<1$ the series 
\begin{multline*}
\sum_{n=1}^{\infty }(-i)^{n}\langle \varphi _{1}\otimes \underset{i\in
S_{k_{-}}}{\prod\nolimits^{\prime }}A^{+}(\tilde{f}_{i}^{-}(\lambda ))\Phi
\;|\;D_{n}(t,\lambda )\varphi _{2}\otimes \prod_{i\in S_{k_{+}}}A^{+}(\tilde{%
f}_{i}^{+}(\lambda ))\Phi \rangle \\
=\langle \phi _{1}\otimes \underset{i\in S_{k_{-}}}{\prod\nolimits^{\prime }}%
A^{+}(\tilde{f}_{i}^{-}(\lambda ))\Phi \;|\;U(t/\lambda ^{2},\lambda )\phi
_{2}\otimes \prod_{i\in S_{k_{+}}}A^{+}(\tilde{f}_{i}^{+}(\lambda ))\Phi
\rangle
\end{multline*}
converges uniformly and absolutely in the pair $(\lambda ,t)\in \mathbb{R}%
_{+}\times \lbrack 0,T]$ for any $T<\infty $. This means that we can pass to
limit under the summation and change the order of summation.

Let us calculate the limit of right part of equation (\ref{n_term}) as $%
\lambda \rightarrow 0$. It follows from lemma 4.3 that only type $I$ terms
will survive under this limit. The equality 
\begin{multline*}
\lim_{\lambda \rightarrow 0}\langle \underset{i\in \bar{P_{1}}|_{S_{k_{-}}}}{%
\prod\nolimits^{\prime }}A^{+}(\tilde{f}_{i}^{-}(\lambda ))\Phi
\;|\;\prod_{i\in \bar{P_{2}}|_{S_{k_{+}}}}A^{+}(\tilde{f}_{i}^{+}(\lambda
))\Phi \rangle = \\
\langle \underset{i\in \bar{P_{1}}|_{S_{k_{-}}}}{\prod\nolimits^{\prime }}%
\mathbb{A}^{+}(f_{i}^{-}\otimes 1_{[S_{i}^{-},T_{i}^{-}]})\Psi
\;|\;\prod_{i\in \bar{P_{2}}|_{S_{k_{+}}}}\mathbb{A}^{+}(f_{i}^{+}\otimes
1_{[S_{i}^{+},T_{i}^{+}]})\Psi \rangle
\end{multline*}
is the immediate consequence of (\ref{f_lim}). The equality $\lim_{\lambda
\rightarrow 0}h_{i}^{\pm }(t,\lambda )=h_{i}^{\pm }(t)$ is due to definition
(\ref{h_lim}). The limit of $G_{\lambda }$ is given by (\ref{G_lim}), but,
since we integrate over the simplex, only the future part $\kappa \frak{d}%
_{+}(u)$ survives. At last, we change the order of summation to $\sum_{%
\mathcal{G}\in \frak{G}_{I,n}}\sum_{\alpha ,\beta \in \{0,1\}^{n}}$ and
replace $\sum_{n=1}^{\infty }\sum_{\mathcal{G}\in \frak{G}_{I,n}}$ by $\sum_{%
\mathcal{G}\in \frak{G}_{I}}$.
\end{proof}

\bigskip

Our next goal is to prove the convergence in the sense of matrix elements of 
$U(t/\lambda ^{2},\lambda )$ to the solution $U_{t}$ of some QSDE. It is
hard work to obtain the explicit form of this equation from (6.1).
Fortunately, we already have this equation for Bosonic case \cite{Gough1}.
The Bosonic version of (6.1) which differs only in the absence of the
Fermionic signs. We can readily predict that QSDE will take the same form
again in the Fermionic case; we shall show below that this indeed is the
case.

Suppose $||\kappa E_{11}||<1$, for $\alpha ,\beta \in \{0,1\}$ we define the
operators $L_{\alpha \beta }:\,\frak{H}_{S}\rightarrow \,\frak{H}_{S}$ by
the rule 
\begin{equation}
L_{\alpha \beta }:=-iE_{\alpha \beta }-\kappa E_{\alpha 1}(1+i\kappa
E_{11})^{-1}E_{1\beta }  \tag{6.2}  \label{L_def}
\end{equation}

By $\frak{G}_{I}(N)$ denote a set of type $I$ partitions $\mathcal{G}$ such
that $N(\mathcal{G})=N$. Clearly, $\frak{G}_{I}=\cup _{N=1}^{\infty }\frak{G}%
_{I}(N)$.

\bigskip

\noindent \textbf{Lemma 6.2: }For any $\alpha ,\beta \in \{0,1\}^{N}$ we
have 
\begin{equation}
L_{\alpha _{n}\beta _{n}}\ldots L_{\alpha _{1}\beta _{1}}=\sum_{\mathcal{G}%
=\left\{ G_{n}>\dots >G_{1}\right\} \in \frak{G}_{I}}^{N\left( \mathcal{G}%
\right) =n}L_{\alpha _{n}\beta _{n}}\left( r_{G_{n}}\right) \ldots L_{\alpha
_{1}\beta _{1}}\left( r_{G_{1}}\right)  \tag{6.3}  \label{resum}
\end{equation}
where parts $G_{1},\dots ,G_{n}\in \mathcal{G}$ are labelled in the obvious
way ($G_{j}>G_{i}$ if all the elements of $G_{j}$ are greater than those of $%
G_{i}$ - as the partition is type $I$ we must either have $G_{j}>G_{i}$ or $%
G_{j}<G_{i}$ for different parts), $r_{G}$ gives the size of a part $G$, and
we introduce the operators $L_{\alpha \beta }\left( r\right) $ 
\begin{equation*}
L_{\alpha \beta }\left( r\right) :=\left\{ 
\begin{array}{cc}
-iE_{\alpha \beta }, & r=1; \\ 
-\kappa E_{\alpha 1}(-i\kappa E_{11})^{r-2}E_{1\beta }, & r>1.
\end{array}
\right.
\end{equation*}
(Note that the summation in (6.3) is over all type $I$ partitions having $n$
parts.)

\begin{proof}
Using definition (\ref{L_def}) of $L_{\alpha \beta }$ we can write 
\begin{eqnarray*}
L_{\alpha _{n}\beta _{n}}\ldots L_{\alpha _{1}\beta _{1}} &=&[-iE_{\alpha
_{n}\beta _{n}}-\kappa \sum_{r_{n}=2}^{\infty }E_{\alpha _{n}1}(-i\kappa
E_{11})^{r_{n}-2}E_{1\beta _{n}}]  \notag \\
&&\times \ldots \times \lbrack -iE_{\alpha _{1}\beta _{1}}-\kappa
\sum_{r_{1}=2}^{\infty }E_{\alpha _{1}1}(-i\kappa E_{11})^{r_{1}-2}E_{1\beta
_{1}}]  \notag \\
&=&\sum_{r_{1},\ldots ,r_{n}=1}^{\infty }L_{\alpha _{n}\beta _{n}}\left(
r_{n}\right) \cdots L_{\alpha _{1}\beta _{1}}\left( r_{1}\right)
\end{eqnarray*}
\begin{equation*}
\tag{6.4}
\end{equation*}
All the series converge absolutely so we can multiply them term by term.
Taking into account that there exists an obvious one-to-one correspondence
between all sets $\{r_{1},\ldots ,r_{n}\}$ and the partitions of $\frak{G}%
_{I}$ having $n$ parts, we see that (\ref{resum}) and (6.4) coincide.
\end{proof}

\bigskip

Define the following four operator processes $\mathbb{A}_{t}^{10}=\mathbb{A}%
^{+}(g\otimes 1_{[0,t]})$, $\mathbb{A}_{t}^{01}=\mathbb{A}^{-}(g\otimes
1_{[0,t]})$, $\mathbb{A}_{t}^{11}=\mathbf{\Lambda }(P_{g}\otimes \chi
_{\lbrack 0,t]})$, $\mathbb{A}_{t}^{00}=t$.

\noindent \textbf{Lemma 6.3:} \label{lem_ad} For any $k_{+},k_{-}\in \mathbb{%
N}_{0}$, $f_{1}^{-},\ldots ,f_{k_{-}}^{-},f_{1}^{+},\ldots ,f_{k_{+}}^{+}\in 
\frak{k}$, $\{S_{i}^{-},T_{i}^{-}\}_{i=1}^{k_{-}}$, $\{S_{i}^{+},T_{i}^{+}%
\}_{i=1}^{k_{+}}\subset \mathbb{R}$, $t\geq 0$, we have 
\begin{multline}  \label{mult_ad}
\langle \underset{i\in S_{k_{-}}}{\prod\nolimits^{\prime }}\mathbb{A}%
^{+}(f_{i}^{-}\otimes 1_{[S_{i}^{-},T_{i}^{-}]})\Psi \;|\; d\mathbb{A}%
_{s_{n}}^{\alpha _{n}\beta _{n}}\ldots d\mathbb{A}_{s_{1}}^{\alpha _{1}\beta
_{1}}\prod_{i\in S_{k_{+}}}\mathbb{A}^{+}(f_{i}^{+}\otimes
1_{[S_{i}^{+},T_{i}^{+}]})\Psi \rangle \\
= \sum_{P_{1}\in \mathcal{P}(S_{k_{-}},S_{n})}\sum_{P_{2}\in \mathcal{P}%
(S_{k_{+}},S_{n})}(-1)^{\xi (\mathcal{G}_{n}^{0},\underline{\alpha },%
\underline{\beta })+\xi (P_{1},\underline{\alpha },1)+\xi (P_{2},1,%
\underline{\beta })} \\
\times \delta _{0}\left( \sum_{i\in \bar{P_{1}}|_{S_{n}}}\alpha
_{i}+\sum_{i\in \bar{P_{2}}|_{S_{n}}}\beta _{i}\right)
\prod_{(i_{1},j_{1})\in P_{1}}\alpha
_{j_{1}}h_{i_{1}}^{-}(s_{j_{1}})\prod_{(i_{2},j_{2})\in P_{1}}(\beta
_{j_{2}}h_{i_{2}}^{+}(s_{j_{2}}))^{\ast } \\
\times \langle \underset{i\in \bar{P_{1}}|_{S_{k_{-}}}}{\prod\nolimits^{%
\prime }}\mathbb{A}^{+}(f_{i}^{-}\otimes 1_{[S_{i}^{-},T_{i}^{-}]})\Psi
\;|\;\prod_{i\in \bar{P_{2}}|_{S_{k_{+}}}}\mathbb{A}^{+}(f_{i}^{+}\otimes
1_{[S_{i}^{+},T_{i}^{+}]})\Psi \rangle,  \tag{6.5}
\end{multline}
where $\mathcal{G}_{n}^{0}$ is the partition of $S_{n}$ which consists of
singletons only.

\begin{proof}
The operator $d\mathbb{A}_{t}^{\alpha\beta }$ satisfy the following
commutational relations 
\begin{equation}  \label{comrel}
d\mathbb{A}_{t}^{\alpha\beta } \mathbb{A}^+(f\otimes 1_{[S,T]}) =
(-1)^{\alpha+\beta}\mathbb{A}^+(f\otimes 1_{[S,T]})d\mathbb{A}%
_{t}^{\alpha\beta } + \beta \langle g\,|\,f \rangle 1_{[S,T]}(t)d\mathbb{A}%
_{t}^{\alpha 0}.  \tag{6.6}
\end{equation}
It is easy to see that the operator $d\mathbb{A}_{t}^{\alpha\beta }$ acts
here as a product $[d\mathbb{A}_{t}^+]^{\alpha}[d\mathbb{A}_{t}^-]^{\beta}$.
Using this fact we can informally justify the equality (\ref{mult_ad}) as
follows.

We replace all $d\mathbb{A}_{s_{i}}^{\alpha_{i}\beta _{i}}$ by products $[d%
\mathbb{A}_{s_{i}}^+]^{\alpha_{i}}[d\mathbb{A}_{s_{i}}^-]^{\beta_{i}}$.
Using the anticommutational relations we get 
\begin{multline*}
[d\mathbb{A}_{s_{n}}^+]^{\alpha_{n}}[d\mathbb{A}_{s_{n}}^-]^{\beta_{n}}%
\ldots [d\mathbb{A}_{s_{1}}^+]^{\alpha_{1}}[d\mathbb{A}_{s_{1}}^-]^{%
\beta_{1}} \\
=(-1)^{\xi (\mathcal{G}_{n}^{0},\underline{\alpha },\underline{ \beta })}[d%
\mathbb{A}_{s_{n}}^+]^{\alpha_{n}}\ldots[d\mathbb{A}_{s_{1}}^+]^{\alpha_{1}}
[d\mathbb{A}_{s_{n}}^-]^{\beta_{n}}\ldots [d\mathbb{A}_{s_{1}}^-]^{%
\beta_{1}}.
\end{multline*}
To bring the expression $[d\mathbb{A}_{s_{n}}^-]^{\beta_{n}}\ldots [d\mathbb{%
A}_{s_{1}}^-]^{\beta_{1}} \prod_{i\in S_{k_{+}}}\mathbb{A}%
^{+}(f_{i}^{+}\otimes 1_{[S_{i}^{+},T_{i}^{+}]})$ to normal order we can use
the equality (\ref{WO_2}). Taking into account that in this case the inner
product $\langle \beta _{j}g_{j}\,|\,\alpha _{i}f_{i}\rangle$ becomes $%
\langle \beta _{j}g \,|\,f^+_i \rangle 1_{[S^+_i,T^+_i]}(s_j)$, which is
exactly $(\beta _{j} h^+_i(s_j))^*$, we obtain (\ref{mult_ad}).

The formal proof are very similar. Note, first of all, that the
representation of $d\mathbb{A}_{t}^{\alpha\beta}$ as a product is justified
when $\alpha$, or $\beta$, or both equals to zero. We only need to show that
we can also factorize $d\mathbb{A}_{t}^{1 1}$.

To obtain the equality (\ref{mult_ad}) from 
\begin{equation*}
\langle \Psi \;|\; \underset{i\in S_{k_{-}}}{\prod}\mathbb{A}%
^{-}(f_{i}^{-}\otimes 1_{[S_{i}^{-},T_{i}^{-}]}) d\mathbb{A}_{s_{n}}^{\alpha
_{n}\beta _{n}}\ldots d\mathbb{A}_{s_{1}}^{\alpha _{1}\beta _{1}}\prod_{i\in
S_{k_{+}}}\mathbb{A}^{+}(f_{i}^{+}\otimes
1_{[S_{i}^{+},T_{i}^{+}]})\Psi\rangle
\end{equation*}
we need to move $d\mathbb{A}_{s_{i}}^{\alpha _{i}\beta _{i}}$ to the right
if $\alpha _{i} = 0$ and to the left if $\beta _{i} = 0$ (and keep in place
if both are zero). The only problem is $d\mathbb{A}_{s_{i}}^{1 1}$ which we
have to move to the right and to the left. Consider the term 
\begin{multline*}
C \times \ldots d\mathbb{A}_{s_{i}}^{\alpha _{i}\beta _{i}} [d\mathbb{A}%
_{s_{i-1}}^+]^{\alpha_{i-1}}\ldots [d\mathbb{A}_{s_{1}}^+]^{\alpha_{1}} \\
\times \prod_{i\in I \subset S_{k_{+}}}\mathbb{A}^{+}(f_{i}^{+}\otimes
1_{[S_{i}^{+},T_{i}^{+}]}) [d\mathbb{A}_{s_{j_r}}^-]^{\beta_{j_r}}\ldots [d%
\mathbb{A}_{s_{j_1}}^-]^{\beta_{j_1}},
\end{multline*}
which appears after we have moved all $[d\mathbb{A}_{s_{j}}^-]^{\beta_{j}}$, 
$j = 1,\ldots,i-1$, to the right. Here $C$ is some factor, $I$ is a subset
of $S_{k_{+}}$ with $i-1 - j_r + |I| = k_{+}$. We suppose that $i$ is the
least index with $\alpha_i = \beta_i = 1$. Moving $d\mathbb{A}%
_{s_{i}}^{\alpha _{i}\beta _{i}}$ to the right and using the commutational
relations (\ref{comrel}), we represent this term as a sum of $|I| + 1$
terms; $|I|$ terms contain contraction and one term is without contraction.
The terms with contraction have the form 
\begin{multline*}
C \times \ldots [d\mathbb{A}_{s_{i-1}}^+]^{\alpha_{i-1}}\ldots [d\mathbb{A}%
_{s_{1}}^+]^{\alpha_{1}} \mathbb{A}^{+}(f_{I(1)}^{+}\otimes
1_{[S_{I(1)}^{+},T_{I(1)}^{+}]})\ldots \\
(\beta _{i} h^+_{I(k(i))}(s_i))^* d\mathbb{A}_{s_{i}}^{\alpha _{i} 0}\ldots 
\mathbb{A}^{+}(f_{I(|I|)}^{+}\otimes 1_{[S_{I(|I|)}^{+},T_{I(|I|)}^{+}]}) [d%
\mathbb{A}_{s_{j_r}}^-]^{\beta_{j_r}}\ldots [d\mathbb{A}_{s_{j_1}}^-]^{%
\beta_{j_1}},
\end{multline*}
where $I(k)$ is a $k$-th element of set $I$. Now we have to move the
operator $d\mathbb{A}_{s_{i}}^{\alpha _{i} 0}$ back to the position which
were occupied by $d\mathbb{A}_{s_{i}}^{\alpha _{i}\beta _{i}}$ before.
Operator $d\mathbb{A}_{s_{i}}^{\alpha _{i} 0}$ anticommute with all $d%
\mathbb{A}^+_t$ and $\mathbb{A}^+$ so we get 
\begin{multline*}
(-1)^{k(i)-1 + \alpha_{i-1}+\ldots+\alpha_1} C \times \ldots d\mathbb{A}%
_{s_{i}}^{\alpha _{i} 0}[d\mathbb{A}_{s_{i-1}}^+]^{\alpha_{i-1}}\ldots [d%
\mathbb{A}_{s_{1}}^+]^{\alpha_{1}} \mathbb{A}^{+}(f_{I(1)}^{+} \otimes \\
1_{[S_{I(1)}^{+},T_{I(1)}^{+}]})\ldots (\beta _{i} h^+_{I(k(i))}(s_i))^*
\ldots \mathbb{A}^{+}(f_{I(|I|)}^{+}\otimes
1_{[S_{I(|I|)}^{+},T_{I(|I|)}^{+}]}) [d\mathbb{A}_{s_{j_r}}^-]^{\beta_{j_r}}%
\ldots [d\mathbb{A}_{s_{j_1}}^-]^{\beta_{j_1}}.
\end{multline*}

This is exactly the same result which we would obtain if we initially had $[d%
\mathbb{A}_{s_{i}}^+]^{\alpha_{i}}[d\mathbb{A}_{s_{i}}^-]^{\beta_{i}}$
instead of $d\mathbb{A}_{s_{i}}^{\alpha _{i}\beta _{i}}$ and moved only the $%
[d\mathbb{A}_{s_{i}}^-]^{\beta_{i}}$ to the right. We can not bring the term
without contraction to the same form but actually we do not need to do this
because the vacuum expectation of this term equals to zero and so its exact
form is of no importance. Thus, we can suppose again for generality that we
had $[d\mathbb{A}_{s_{i}}^+]^{\alpha_{i}}[d\mathbb{A}_{s_{i}}^-]^{\beta_{i}}$
instead of $d\mathbb{A}_{s_{i}}^{\alpha _{i}\beta _{i}}$.

Now each of new terms has the same form as the initial one and we can repeat
the above procedure for the next $d\mathbb{A}_{s}^{1 1}$. This proves lemma.
\end{proof}

\bigskip

\noindent \textbf{Theorem 6.4:}\label{QSDE} Suppose $K||E_{11}||<1$, then
for any $\varphi _{1},\varphi _{2}\in \,\frak{H}_{S}$, $k_{+},k_{-}\in 
\mathbb{N}_{0}$, $f_{1}^{-},\ldots ,f_{k_{-}}^{-},f_{1}^{+},\ldots
,f_{k_{+}}^{+}\in \frak{k}$, $\{S_{i}^{-},T_{i}^{-}\}_{i=1}^{k_{-}},%
\{S_{i}^{+},T_{i}^{+}\}_{i=1}^{k_{+}}\subset \mathbb{R}$, $t\geq 0$, we have 
\begin{multline}
\lim_{\lambda \rightarrow 0}\langle \phi _{1}\otimes \underset{i\in S_{k_{-}}%
}{\prod\nolimits^{\prime }} A^{+}(\tilde{f}_{i}^{-}(\lambda ))\Phi
\;|\;U(t/\lambda ^{2},\lambda )\,\phi _{2}\otimes \prod_{i\in
S_{k_{+}}}A^{+}(\tilde{f}_{i}^{+}(\lambda ))\Phi \rangle =  \notag \\
\langle \phi _{1}\otimes \underset{i\in S_{k_{-}}}{\prod\nolimits^{\prime }} 
\mathbb{A}^{+}(f_{i}^{-}\otimes 1_{[S_{i}^{-},T_{i}^{-}]})\Psi
\;|\;U_{t}\,\phi _{2}\otimes \prod_{i\in S_{k_{+}}}\mathbb{A}%
^{+}(f_{i}^{+}\otimes 1_{[S_{i}^{+},T_{i}^{+}]})\Psi \rangle ,  \tag{6.7}
\end{multline}
where $(U_{t})_{t}$ is a unitary adapted quantum stochastic process on $\,%
\frak{H}_{S}\otimes \Gamma _{-}(\frak{k}\otimes L^{2}(\mathbb{R}^{+}))$
satisfying the quantum stochastic differential equation 
\begin{equation}
dU_{t}=L_{\alpha \beta }U_{t} \otimes d\mathbb{A}_{t}^{\alpha \beta } 
\tag{6.8}
\end{equation}
with $U_{0}=1$ and $L_{\alpha \beta }$ given by (\ref{L_def}).

\begin{proof}
We have the following expansion for $U_{t}$: 
\begin{equation}
U_{t}=\sum_{n=0}^{\infty }\sum_{\alpha ,\beta \in
\{0,1\}^{n}}\int_{\triangle _{n}(t)}L_{\alpha _{n}\beta _{n}}\ldots
L_{\alpha _{1}\beta _{1}}\otimes d\mathbb{A}_{s_{n}}^{\alpha _{n}\beta
_{n}}\ldots d\mathbb{A}_{s_{1}}^{\alpha _{1}\beta _{1}}.  \tag{6.9}
\end{equation}

From (\ref{resum}) and (\ref{mult_ad}) we obtain 
\begin{multline}
\langle \phi _{1}\otimes \underset{i\in S_{k_{-}}}{\prod\nolimits^{\prime }}%
\mathbb{A}^{+}(f_{i}^{-}\otimes 1_{[S_{i}^{-},T_{i}^{-}]})\Psi
\;|\;U_{t}\,\phi _{2}\otimes \prod_{i\in S_{k_{+}}}\mathbb{A}%
^{+}(f_{i}^{+}\otimes 1_{[S_{i}^{+},T_{i}^{+}]})\Psi \rangle \\
=\sum_{n=0}^{\infty }\sum_{\alpha ,\beta \in \{0,1\}^{n}}\sum_{\mathcal{G}%
=\left\{ G_{n}>\dots >G_{1}\right\} \in \frak{G}_{I}}^{N\left( \mathcal{G}%
\right) =n}\langle \varphi _{1}\;|\;L_{\alpha _{n}\beta _{n}}\left(
r_{G_{n}}\right) \ldots L_{\alpha _{1}\beta _{1}}\left( r_{G_{1}}\right)
\varphi _{2}\rangle  \notag \\
\times \sum_{P_{1}\in \mathcal{P}(S_{k_{-}},S_{n})}\sum_{P_{2}\in \mathcal{P}%
(S_{k_{+}},S_{n})}(-1)^{\xi (\mathcal{G}_{n}^{0},\underline{\alpha },%
\underline{\beta })+\xi (P_{1},\underline{\alpha },1)+\xi (P_{2},1,%
\underline{\beta })}\delta _{0}\left( \sum_{i\in \bar{P_{1}}|_{S_{n}}}\alpha
_{i}\right.  \notag \\
\left.+\sum_{i\in \bar{P_{2}}|_{S_{n}}}\beta _{i}\right) \langle \underset{%
i\in \bar{P_{1}}|_{S_{k_{-}}}}{\prod\nolimits^{\prime }}\mathbb{A}%
^{+}(f_{i}^{-}\otimes 1_{[S_{i}^{-},T_{i}^{-}]})\Psi \;|\;\prod_{i\in \bar{%
P_{2}}|_{S_{k_{+}}}}\mathbb{A}^{+}(f_{i}^{+}\otimes
1_{[S_{i}^{+},T_{i}^{+}]})\Psi \rangle  \notag \\
\times \int_{\triangle _{n}(t)}ds_{n}\ldots ds_{1}\prod_{(i_{1},j_{1})\in
P_{1}}\alpha _{j_{1}}h_{i_{1}}^{-}(s_{j_{1}})\prod_{(i_{2},j_{2})\in
P_{1}}(\beta _{j_{2}}h_{i_{2}}^{+}(s_{j_{2}}))^{\ast }.  \tag{6.10}
\end{multline}
Now we make a slight change of notation. Suppose $\mathcal{G}\in \frak{G}%
_{I} $ consists of parts with sizes $r_{1},\ldots ,r_{n}$. The $j$th part
will have $q_{j}=1+\sum_{h<j}r_{h}$ as minimum element and $%
p_{j}=\sum_{h\leq j}r_{h}$ as maximum element. Given $\underline{\alpha },%
\underline{\beta }\in \left\{ 0,1\right\} ^{n}$ we define $\underline{\hat{%
\alpha}\left( \mathcal{G}\right) },\underline{\hat{\beta}\left( \mathcal{G}%
\right) }\in \left\{ 0,1\right\} ^{E\left( \mathcal{G}\right) }$ by $\hat{%
\alpha}_{j}\left( \mathcal{G}\right) =\hat{\beta}_{j}\left( \mathcal{G}%
\right) =1$ with the exceptions 
\begin{equation}
\hat{\alpha}_{p_{j}}(\mathcal{G},\alpha )=\alpha _{j};\;\;\hat{\beta}%
_{q_{j}}(\mathcal{G},\alpha )=\beta _{j},  \tag{6.11}
\end{equation}
Note that $\{q_{j}:j=1,\dots ,n\}=\mathcal{G}_{in}$ and $\{p_{j}:j=1,\dots
,n\}=\mathcal{G}_{out}$. We then can write 
\begin{eqnarray*}
&&\sum_{\alpha ,\beta \in \{0,1\}^{n}}\sum_{\mathcal{G}=\left\{ G_{n}>\dots
>G_{1}\right\} \in \frak{G}_{I}}^{N\left( \mathcal{G}\right) =n}\langle
\varphi _{1}\;|\;L_{\alpha _{n}\beta _{n}}\left( r_{G_{n}}\right) \ldots
L_{\alpha _{1}\beta _{1}}\left( r_{G_{1}}\right) \varphi _{2}\rangle  \notag
\\
&=&\sum_{\mathcal{G}\in \frak{G}_{I}}^{N\left( \mathcal{G}\right)
=n}\sum_{\alpha ,\beta \in \{0,1\}^{n}}(-i)^{E(\mathcal{G})}\kappa ^{E(%
\mathcal{G})-n}\langle \varphi _{1}\;|\;E_{\hat{\alpha}_{E(\mathcal{G})}(%
\mathcal{G})\hat{\beta}_{E(\mathcal{G})}(\mathcal{G})}\ldots E_{\hat{\alpha}%
_{1}(\mathcal{G})\hat{\beta}_{1}(\mathcal{G})}\varphi _{2}\rangle  \notag \\
&=&\sum_{\mathcal{G}\in \frak{G}_{I}}^{N\left( \mathcal{G}\right)
=n}\sum_{\alpha ,\beta \in \{0,1\}^{n}}(-i)^{E(\mathcal{G})}\langle \varphi
_{1}\;|\;E_{\hat{\alpha}_{E(\mathcal{G})}(\mathcal{G})\hat{\beta}_{E(%
\mathcal{G})}(\mathcal{G})}\ldots E_{\hat{\alpha}_{1}(\mathcal{G})\hat{\beta}%
_{1}(\mathcal{G})} \varphi _{2}\rangle \\
&\times&\!\!\!\!\! \prod_{\{i(r)>\ldots >i(1)\}\in \mathcal{G}}\kappa ^{r-1}%
\hat{\beta}_{i(r)}(\mathcal{G})\hat{\alpha}_{i(r-1)}(\mathcal{G})\hat{\beta}%
_{i(r-1)}(\mathcal{G})\ldots \hat{\alpha}_{i(2)}(\mathcal{G})\hat{\beta}%
_{i(2)}(\mathcal{G})\hat{\alpha}_{i(1)}(\mathcal{G}) \\
&=&\sum_{\mathcal{G}\in \frak{G}_{I}}^{N\left( \mathcal{G}\right) =n}\sum_{%
\hat{\alpha} ,\hat{\beta} \in \{0,1\}^{E(\mathcal{G})}}(-i)^{E(\mathcal{G}%
)}\langle \varphi _{1}\;|\;E_{\hat{\alpha}_{E(\mathcal{G})}\hat{\beta}_{E(%
\mathcal{G})}}\ldots E_{\hat{\alpha}_{1}\hat{\beta}_{1}} \varphi _{2}\rangle
\\
&&\times \prod_{\{i(r)>\ldots >i(1)\}\in \mathcal{G}}\kappa ^{r-1}\hat{\beta}%
_{i(r)}\hat{\alpha}_{i(r-1)}\hat{\beta}_{i(r-1)}\ldots \hat{\alpha}_{i(2)}%
\hat{\beta}_{i(2)}\hat{\alpha}_{i(1)}
\end{eqnarray*}
The multiplication by $\prod_{\{i(r)>\ldots >i(1)\}\in \mathcal{G}}\hat{\beta%
}_{i(r)}(\mathcal{G})\ldots \hat{\alpha}_{i(1)}(\mathcal{G})$ does not
affect the equality because it equals to unity; we are then able to extend
the summation over all $\underline{\hat{\alpha}},\underline{\hat{\beta}}\in
\{0,1\}^{E(\mathcal{G})}$ because this multiplier makes all additional terms
equal to zero.

Now we are going to rewrite (6.9) in terms of extended parameters $%
\underline{\hat{\alpha}},\underline{\hat{\beta}}$, which appeared from the
original ones $\underline{\alpha },\underline{\beta }$ due to a given $%
\mathcal{G}\in \frak{G}_{I}$. To do this we only need to ``restore the
indices'' to $\underline{\alpha },\underline{\beta }$ in $\underline{\hat{%
\alpha}},\underline{\hat{\beta}}$, that is, we have to replace $P_{1}\in 
\mathcal{P}(S_{k_{-}},S_{n})$ by $\hat{P}_{1}\in \mathcal{P}(S_{k_{-}},%
\mathcal{G}_{out})$ and replace $P_{2}\in \mathcal{P}(S_{k_{+}},S_{n})$ by $%
\hat{P}_{2}\in \mathcal{P}(S_{k_{+}},\mathcal{G}_{in})$. Clearly, 
\begin{equation*}
\delta _{0}\left( \sum_{i\in \Bar{\Hat{P}}_{1}|_{\mathcal{G}_{out}}}\hat{%
\alpha}_{i}+\sum_{i\in \Bar{\Hat{P}}_{2}|_{\mathcal{G}_{in}}}\hat{\beta}%
_{i}\right) =\delta _{0}\left( \sum_{i\in \bar{P_{1}}|_{S_{n}}}\alpha
_{i}+\sum_{i\in \bar{P_{2}}|_{S_{n}}}\beta _{i}\right) .
\end{equation*}

It is easy to check that $\xi (\mathcal{G}_{n}^{0},\alpha ,\beta )=\xi (%
\mathcal{G},\hat{\alpha}(\mathcal{G}),\hat{\beta}(\mathcal{G}))$ for all $%
\mathcal{G}\in \frak{G}_{I}$ such that $N(\mathcal{G}) = n$, $\xi (\hat{P}%
_{1},\hat{\alpha}(\mathcal{G}),1)=\xi (P_{1},\alpha ,1)$ and $\xi (\hat{P}%
_{2},1,\hat{\beta}(\mathcal{G}))=\xi (P_{2},1,\beta )$.

The subintegral expression we shall write as follows: First, we rename the
integral variables to $s_{q_{1}},s_{q_{2}},\ldots ,s_{q_{n}}$. Then we
extend the integration to $\triangle _{E(\mathcal{G})}$ by adding the $\frak{%
d}_{+}$-functions $\prod_{\{i(r)>\ldots >i(1)\}\in \mathcal{G}%
}\prod_{h=1}^{r-1}\frak{d}_{+}(s_{i\left( h+1\right) }-s_{i\left( h\right) })
$ where we multiply out over all parts of $\mathcal{G}$. Combining all
together finally we get (we now omit the ``hats'') 
\begin{multline}  \label{equ612}
\langle \phi _{1}\otimes \underset{i\in S_{k_{-}}}{\prod\nolimits^{\prime }}%
\mathbb{A}^{+}(f_{i}^{-}\otimes 1_{[S_{i}^{-},T_{i}^{-}]})\Psi
\;|\;U_{t}\phi _{2}\otimes \prod_{i\in S_{k_{+}}}\mathbb{A}%
^{+}(f_{i}^{+}\otimes 1_{[S_{i}^{+},T_{i}^{+}]})\Psi \rangle  \notag \\
=\sum_{\mathcal{G}\in \frak{G}_{I}}\sum_{\alpha ,\beta \in \{0,1\}^{E(%
\mathcal{G})}}\langle \varphi _{1}\,|\,E_{\alpha _{E(\mathcal{G})}\beta _{E(%
\mathcal{G})}}\ldots E_{\alpha _{1}\beta _{1}}\varphi _{2}\rangle
\sum_{P_{1}\in \mathcal{P}(S_{k_{-}},\mathcal{G}_{out})}\sum_{P_{2}\in 
\mathcal{P}(S_{k_{+}},\mathcal{G}_{in})}  \notag \\
(-i)^{E(\mathcal{G})}(-1)^{\xi (\mathcal{G},\alpha ,\beta )+\xi
(P_{1},\alpha ,1)+\xi (P_{2},1,\beta )}\delta _{0}\left( \sum_{i\in \bar{%
P_{1}}|_{\mathcal{G}_{out}}}\alpha _{i}+\sum_{i\in \bar{P_{2}}|_{\mathcal{G}%
_{in}}}\beta _{i}\right)  \notag \\
\times \langle \prod_{i\in \bar{P_{1}}|_{S_{k_{-}}}}\mathbb{A}%
^{+}(f_{i}^{-}\otimes 1_{[S_{i}^{-},T_{i}^{-}]})\Psi \;|\;\prod_{i\in \bar{%
P_{2}}|_{S_{k_{+}}}}\mathbb{A}^{+}(f_{i}^{+}\otimes
1_{[S_{i}^{+},T_{i}^{+}]})\Psi \rangle  \notag \\
\times \int_{\triangle _{E\left( \mathcal{G}\right) }(t)}ds_{n(\mathcal{G}%
)}\ldots ds_{1}\prod_{(i_{1},j_{1})\in P_{1}}\alpha
_{j_{1}}h_{i_{1}}^{-}(s_{j_{1}})\prod_{(i_{2},j_{2})\in P_{1}}(\beta
_{j_{2}}h_{i_{2}}^{+}(s_{j_{2}}))^{\ast }  \notag \\
\times \prod_{\{i(r)>\ldots >i(1)\}\in \mathcal{G}}\kappa
^{r-1}\prod_{h=1}^{r-1}\left\{ \beta _{i(h+1)}\alpha _{i(h)}\frak{d}%
_{+}(s_{i\left( h+1\right) }-s_{i\left( h\right) })\right\}.  \tag{6.12}
\end{multline}
The right sides of equalities (\ref{equ612}) and (\ref{equ61}) are the same
and this completes the proof.
\end{proof}


\bigskip

\begin{thebibliography}{9}
\bibitem{Applebaum-Hudson}  Applebaum, D., Hudson, R.L.: Fermion Ito's
formula and stochastic evolutions, Commun. Math. Phys. 96, 473-496 (1984)

\bibitem{Applebaum}  Applebaum, D.: Fermion Ito's formula II: the gauge
process in Fermion Fock space, Publ. RIMS,\ Kyoto University 23, 17-56 (1987)

\bibitem{Gough1}  Gough, J.: Quantum Markovian approximation as a quantum
central limit. Preprint, Mathematics and Statistics Research Report Series
No. 11/03, Nottingham-Trent University.

\bibitem{ACFRLU}  Accardi, L., Frigerio, L., Lu, Y.G.: The weak coupling
limit for Fermions. J. Math. Phys. 32 (6), June, 1567-1581 (1991)

\bibitem{HP:UFBSC}  Hudson, R.L., Parthasarathy, K.R.: Unification of
Fermion and Boson stochastic calculus. Commun.Math.Phys. \textbf{104},
457-470 (1986)

\bibitem{Partha}  Parthasarathy, K.R.: Introduction to Quantum Stochastic
Calculus. Basel, Birkh\"{a}user (1992)
\end{thebibliography}
\end{document}